# START as the detector of choice for large-scale muon triggering systems

A. Akindinov[a,*], G. Bondarenko[b], V. Golovin[c], E. Grigoriev[d], Yu. Grishuk[a],
D. Mal'kevich[a], A. Martemiyanov[a], A. Nedosekin[a], M. Ryabinin[a], V. Sheinkman[a],
A. Smirnitskiy[a] and K. Voloshin[a]

[a]*Institute for Theoretical and Experimental Physics (ITEP),
B. Cheremushkinskaya 25, Moscow 117218, Russia*
[b]*Moscow Engineering and Physics Institute,
Kashirskoe shosse 31, Moscow 115409, Russia*
[c]*Center of Perspective Technologies and Apparatus (CPTA),
Preobrazhenskaya pl. 6/8, Moscow 107076, Russia*
[d]*University of Geneva, CMU, rue Michel-Servet 1, Genève 4, 1211, Switzerland*

**Abstract**

Further progress in building high-granular large-scale systems based on Scintillation Tiles with MRS APD light readout (START) became possible thanks to the creation of an improved version of MRS APD. The cost of the system may now be significantly reduced by using inexpensive extruded scintillator. More than 160 START samples were assembled based on this design modification and proved to possess 100% MIP detection efficiency and the intrinsic noise rate of less than 0.08 Hz. Long-term stability of START characteristics was confirmed after 3.5 months of operation.

*Key words:* Scintillation tile, Avalanche photodiode, quantum efficiency, photo-ionization coefficient, MRS APD, WLS fiber, extruded scintillator, ALICE TOF

*PACS:* 07.60.Vg, 29.40.Mc, 29.40.Wk, 81.05.Lg, 85.60.Dw

## 1. Introduction

Avalanche photo-diode, represented by a matrix of metal-resistor-semiconductor micro-cells and operated in the Geiger mode, has long since become an attractive device in high-energy physics instrumentation. MRS APD [1], a successful realization of this technique, has been under development at the Center of Perspective Technologies and Apparatus (CPTA) since the early 1990's. The first encouraging results on MRS APD, reported at the Beaune conference nine years ago [2], launched intensive research on employing this photo-diode for the light readout in scintillating tile-fiber systems. Such systems seemed to have good prospects in muon triggering, calorimetry, medical and astrophysical applications. Within the next several years these efforts led to a significant increase in the quantum efficiency of MRS APD for the visible light and to a reduction of its photo-ionization coefficient [3].

---

[*] Corresponding author.
*Email address:* `Alexandre.Akindinov@cern.ch` (A. Akindinov).

Advantages of MRS APD over traditional multi-anode photo-multiplying tubes include:

(1) miniature dimensions with the sensitive area on the order of 1 mm$^2$ (which allows installing MRS APD directly inside scintillating plastic)
(2) low and safe bias voltage of 20–50 V (depending on the photo-diode version, see Section 2)
(3) low power consumption with the typical supply current in the range of 1–5 μA
(4) no need for special housing
(5) resemblance of main characteristics of photo-diodes produced in lots (which facilitates the construction of large arrays of MRS APDs)
(6) absence of cross-talks between tightly installed photo-diodes
(7) insensitivity to magnetic field [4]
(8) radiation hardness [5]
(9) low cost

START, a Scintillation Tile with MRS APD Light Readout [6], was initially developed for the purpose of building a high-granular and efficient cosmic muon triggering system. Its characteristics depend on the parameters of its main constituents, viz. a pair of MRS APDs, a scintillating plastic plate and a piece of wavelength-shifting (WLS) optical fiber embedded into the plastic. This article reports on the recent progress made in the improvement of these parameters and — consequently — in START performance. Special emphasis is placed on the potential of START cost reduction, which is essential for mass production.

## 2. Advances in MRS APD

The most recent MRS APD samples, produced by CPTA at the end of 2004, have been thoroughly studied at ITEP. Fig. 1 shows a photograph of the MRS APD sensitive surface taken through a microscope ocular. The micro-cell structure of the photo-diode is clearly visible.

The main task for the tests was to measure quantum efficiencies $QE$ and photo-ionization coefficients $\alpha$ of all examined photo-diodes. The latter value is defined as the relative fall of the MRS APD noise counting rate with respect to the increase of the discriminating threshold from the position of one of the photo-electron peaks to the next in the amplitude spectrum. In fact, $\alpha$ characterizes the level of optical cross-talks between neighboring micro-cells in the photo-diode and — altogether with $QE$ — is crucial for the choice of the optimal bias voltage, individually chosen for each MRS APD [1,3]. The basic parameters of MRS APD as well as the average measurement results are summarized in Table 1. As may be seen, the quantum efficiency of MRS APD for the wavelength of $\lambda = 560$ nm exceeds 30%, while the photo-ionization coefficient is as low as 1/15. For comparison, listed alongside are the data obtained with previous versions of MRS APD and corresponding references.

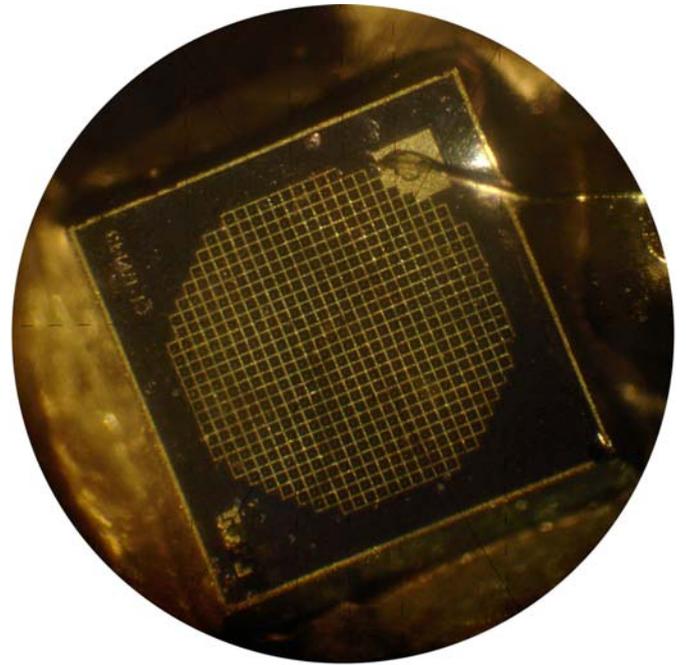

Fig. 1. Microscope view of the sensitive surface of MRS APD.

Table 1. Comparison of different MRS APD versions.

| Parameter | Ref. [2] (1996) | Refs. [1,3] (2001–2003) | New version (2004) |
| --- | --- | --- | --- |
| Size of the sensitive area | $0.5 \times 0.5$ mm$^2$ | $1.0 \times 1.0$ mm$^2$ | $\varnothing 1.15$ mm |
| Number of micro-cells | $\approx 700$ | $\approx 1400$ | 556 |
| Micro-cell size | dot p–n-junction | $20 \times 30$ μm | $44 \times 44$ μm |
| Maximum gain | $5 \cdot 10^5$ | $10^6$ | $3 \cdot 10^6$ |
| Average $\alpha$ | 1/3 | 1/10 | 1/15 |
| Average $QE$ for $\lambda = 560$ nm | < 5% | 20–25% | 30–35% |
| Operating point in the bias voltage | 38–43 V | 48–52 V | 22–27 V |

# 3. START: adaptation for mass production

Scintillation Tile with MRS APD Light Readout (START) [6] has proved to be a reliable and efficient tool for muon triggering [7]. Moreover, it remains a promising instrument for calorimetry, medical and astrophysical imaging. START normally consists of a scintillating plastic plate, a piece of wavelength-shifting (WLS) optical fiber (typically Y11[1]) laid in a circular groove inside the plate, two MRS APDs, firmly pressed to both ends of the fiber and operated in coincidence, an opaque wrapper and a front-end card mounted directly on the detector [6]. Rectangular shape of the plastic plates allows the coverage of unrestrictedly large areas with STARTs, the size of 'blind' zones being limited only by the thickness of the wrapper. A compact and low-maintenance device, START represents a convincing alternative to traditional scintillation counters, in which the light readout is performed by photo-multiplying tubes.

A 32-channel operational muon triggering telescope was assembled of STARTs, sized $15 \times 15 \times 1.5$ cm$^3$, in 2004 [7]. It was developed as a prototype of a much larger cosmic muon facility intended for regular tests of the ALICE TOF system components [8]. The construction of the facility places significant restrictions on the overall cost of its constituent parts, implying that the single START price must decrease dramatically as soon as the mass assemblage begins. This problem is partially addressed by the fact that the cost of wafers for MRS APD is independent of the number of photo-diodes produced in one series, resulting in lower price per unit.

The other way to reduce the cost of START is through using less expensive scintillating plastic. Along with the 1.5 cm-thick BC-412[2] used in the 32-channel prototype [7], a much cheaper extruded scintillator Polisterol-165[3], 1 cm thick (which is a technological limit), has been investigated. Within the START geometry and in conjunction with Y11 WLS fiber, the light yield of BC-412 plates is 1.7 times larger than that of Polisterol-165 plates. However, the use of new MRS APD with higher quantum efficiency may compensate for this ratio. To verify this expectation, two STARTs, both made of Polisterol-165 plates and equipped with the previous (2001–2003, Refs. [1,3]) and latest (2004) versions of MRS APD, were studied with a 1.28 GeV/$c$ $\pi^-$-beam at the ITEP PS. Fig. 2 compares the self-triggered amplitude spectra obtained in each case from one of the two MRS APDs. Sigma-estimation of the number of photo-electrons [2] shows that the scintillation light collected over $15 \times 15 \times 1$ cm$^3$ Polisterol-165 plate produces on average 16 photo-electrons in the modern version of MRS APD (Fig. 2b). This is an excellent result when compared to 11 photo-electrons produced in the older version of the photo-diode (Fig. 2a). Setting the discriminating threshold at the value corresponding to 3.5 photo-electrons, results in 100% MIP detection efficiency by START equipped with modern MRS APDs. The noise rate in this case has been found to be less than 0.05 Hz.

---

[1] By Kuraray Co., Ltd., www.kuraray.co.jp/en/index.html.

[2] By Saint-Gobain Crystals & Detectors, www.bicron.com.

[3] By Polimersintez, Vladimir, Russia.

## 4. START: long term stability and improved performance

Sixteen STARTs with the best performance were selected from the components of the 32-channel cosmic muon system prototype [7] and installed into the ALICE/LHC TOF test facility [8] as shown in the photograph in Fig. 3. Two strips of eight STARTs, intended to pick up cosmic muons, were positioned one over another on the opposite sides of the ALICE TOF module put under tests. After 3.5 months of continuous operation all the 16 STARTs were examined. No performance deterioration has been found and the MIP detection efficiency of all cells remained above 99%.

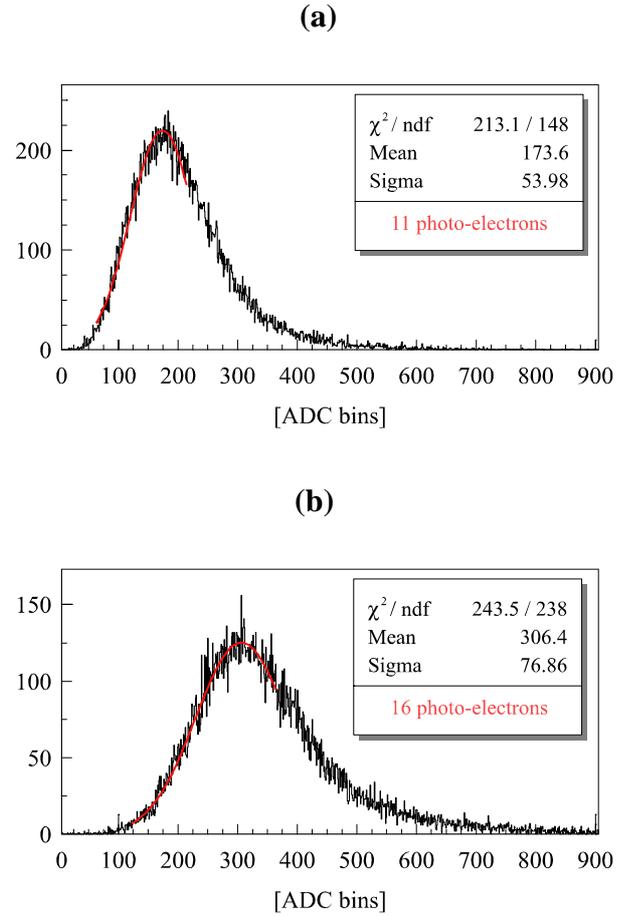

Fig. 2. Comparison of amplitude spectra of two STARTs, assembled of Polisterol-165 plastic plates and equipped with (a) the previous (2001–2003, Refs. [1,3]) and (b) the modern (2004) versions of MRS APD.

The spread of the numbers of photo-electrons between the STARTs in the 32-channel cosmic muon system prototype [7] was found to be larger than expected. This effect may be attributed to the low coupling quality of the fibers to the photo-diodes, which might exist in those detectors. The dimensions and shape of the sensitive area of the new MRS APD were adjusted to allow a better coupling to the WLS fiber edge, resulting in more efficient scintillation light collection and consequently in a smaller spread of the number of photo-electrons, which has been already observed with new STARTs.

More than 160 new START samples were assembled and tested at ITEP and CERN in March-July 2005 in the framework of the ALICE/LHC TOF project. All of them were made of 15 × 15 × 1 cm$^3$ Polisterol-165 plates and equipped with the modern version of MRS APD. The assembled detectors proved to possess MIP detection efficiency of more than 99% and intrinsic noise of less than 0.08 Hz, which corresponds to less than 5% background level in the cosmic muon triggering rate.

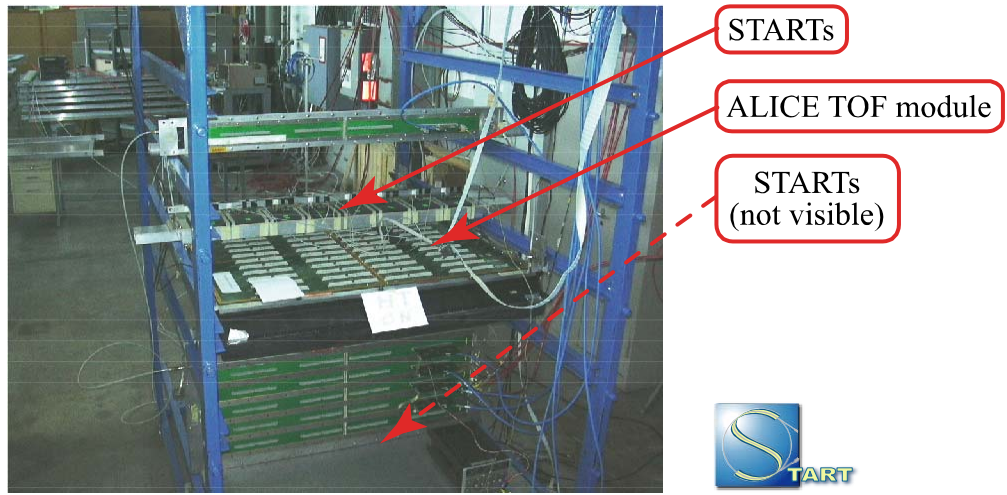

Fig. 3. Prototype of the ALICE TOF cosmic ray facility with 16 STARTs used to trigger cosmic muons.

## 5. Conclusion

Recent achievements in MRS APD development have granted additional advantages to START in terms of its implementation in high-granular, low-noise and efficient muon triggering systems. Cost reduction prospects for such systems are now real due to the possibility of using less expensive extruded scintillator. The reliability of START has been proven in multiple tests and is supported by the evidence of its long-term stable operation. Construction of many START samples is technologically simple and has become an established process.

## References

[1] A. V. Akindinov, G. B. Bondarenko, V. M. Golovin, et al., Instrum. Exp. Tech. 48–3 (2005) 355.

[2] A. V. Akindinov, A. N. Martemianov, P.A. Polozov, et al., Nucl. Instr. and Meth. A 387 (1997) 231.

[3] A. Akindinov, G. Bondarenko, V. Golovin, et al., Proceedings of the 8th ICATPP Conference on Advanced Technology and Particle Physics, Villa Erba, Como, 6-10 October 2003, World Scientific, Singapore, 2004.

[4] D. Beznosko, G. Blazey, A. Dyshkant, et al., *Effects of the Strong Magnetic Field on LED, Extruded Scintillator and MRS Photodiode*, Submitted to Nucl. Instr. and Meth. A.

[5] D. Chakraborty, Nucl. Instr. Meth. A, this issue.

[6] A. Akindinov, G. Bondarenko, V. Golovin, et al., Nucl. Instr. and Meth. A 539 (2005) 172.

[7] A. Akindinov, A. Alici, P. Antonioli, et al., *Prototype of a cosmic muon detection system based on Scintillation counters with MRS APD light readout*, Accepted by Nucl. Instr. and Meth. A.

[8] ALICE Collaboration, Time-of-flight system, Addendum to ALICE TDR 8, CERN/LHCC 2002–16.